# Enhancing gate control and mitigating short channel effects in 20-50 nanometer channel length amorphous oxide Thin-Film Transistors

Chankeun Yoon,[1,2,†] Juhan Ahn,[1,†] Yuchen Zhou,[1,2] Jaydeep P. Kulkarni,[1] and Ananth Dodabalapur[1,2,*]

*[1] Chandra Family Department of Electrical and Computer Engineering, The University of Texas at Austin, Austin, Texas 78712, USA*

*[2]Microelectronics Research Center, The University of Texas at Austin, Austin, Texas 78758, USA*

[†] *These authors equally contributed to this work*

***Email: ananth.dodabalapur@engr.utexas.edu***

**Abstract**

**Field-effect transistors (FETs) with single gates are adversely affected by short channel effects such as drain-induced barrier lowering (DIBL) and increases in the magnitude of sub-threshold swing as the channel length is reduced. Dual-gate and gate-all-around geometries are often employed to improve gate control in very short channel length transistors. This can**

**introduce significant process complexity to the device fabrication compared to single-gate transistors. It is shown in this paper that substantial reductions in short channel effects are possible in single-gate field-effect transistors with indium gallium zinc oxide semiconductor channels by modifying the design of the source and drain electrodes to possess an array of tapered tips which are designated as nanospike electrodes. 20-25 nm channel length FETs with nanospike electrodes have DIBL and other key metrics that are comparable to those in much larger (70-80 nm) channel length FETs with a conventional source/drain electrode design. These improvements stem from better gate control near the source and drain electrode tips due to the shape of these electrodes. These bottom gate FETs had a gate insulator consisting of 9 nm thick $Al_2O_3$ and independent Ni gates. This design approach is expected to be very helpful for a variety of semiconductor technologies being considered for back-end-of-line (BEOL) applications. Simulations with SYNOPSYS SENTAURUS® were performed to understand the device physics of these FETs and facilitate a more detailed comparison**

## Introduction

In all field-effect transistors (FETs), gate control of the channel current, both below and above threshold, becomes increasingly difficult as the channel length is progressively reduced. A set of deleterious effects, known collectively as short channel effects (SCE) start to manifest, requiring a redesign of the device architecture for efficient FET operation. In silicon field-effect transistors, this has resulted in the predominant device architecture changing from the earlier single gate FETs (1,2) to dual gate, trigate or FINFETs (3), and eventually to gate-all-around architectures (4,5). To quantify the relationship between SCE and important FET parameters including channel length, $L_{ch}$, a parameter $\lambda$ was introduced by Yan *et al.* (6), and is currently widely used by the research community. The ratio $L_{ch}/\lambda$ is a key metric in assessing the performance of a FET and typically needs to be more than ~ 5 for the FET to have good operating characteristics above and below threshold (7). Simple algebraic forms for $\lambda$ in terms of channel parameters have been derived for single gate, double gate, as well as some more complex FET architectures (6). In general, the more extensive the electrostatic gate control of channel current, the smaller is the value of $\lambda$ and hence larger $L_{ch}/\lambda$. For single gate FETs, $\lambda$ is given by $\sqrt{(t_{oxide}t_{channel}\varepsilon_{channel})/\varepsilon_{oxide}}$, where $t$ and $\varepsilon$ are thickness and permittivity, respectively. (6,7)

It is shown in this paper that as an alternative approach to the use of double gate or more complex gate architectures, single gate FETs can be modified to result in increased effective $L_{ch}/\lambda$ values for otherwise similar channel parameters. The essence of this approach is a specific change made to the source and drain electrode shape and configuration. As a result of such a change, it is shown that 20 nm channel length FETs with the new design have key SCE parameters such as drain induced barrier lowering (DIBL) that are comparable to those in 50-100 nm channel length FETs with a conventional electrode design that is used by most research groups. This indicates that it is possible to use single gate architectures down to much smaller channel lengths (by a factor of > 2) than is otherwise possible. This can greatly simplify the device architecture, and hence reduce fabricational complexity and cost, and will be especially useful in back-end-of-line (BEOL) circuitry comprised of amorphous oxide thin-film transistors (TFTs) (8-15). Such transistors are being actively considered for memories including dynamic random-access (16-19), artificial intelligence accelerators (20), neuromorphic circuits (21), etc. A BEOL FET technology that is powerful and can be scaled down to very small channel lengths will be useful for designing new types of artificial intelligence (AI) hardware in conjunction with front-end silicon circuitry (22).

The main idea that is demonstrated in this paper is that a source-drain electrode geometry that has spike shaped tips (referred to as nanospikes), as illustrated in **Fig. 1**, results in substantial improvement of key FET parameters such as DIBL and sub-threshold swing (SS). This electrode design is illustrated in **Fig. 1** along with the conventional or flat-edge electrode design that is typically used by most groups. This architecture only requires a modification of the source-drain pattern and does not add any process steps or new materials to the device design. The advantages as well as trade-offs that accrue with this electrode design are discussed in a later section of the paper. In earlier publications, some qualitative aspects of this design were described for larger channel length FETs with unpatterned gates and much thicker (90 nm) $SiO_2$ gate insulators (23-27) and with multilayer $MoS_2$ (28). In this paper, the channel lengths are scaled down to 20-50 nm and the gate insulator equivalent oxide thickness is scaled down to ~4.7 nm. Detailed comparisons are made with equivalent conventional or flat-edge electrode single gate FETs (illustrated schematically in **Fig. 1a**), which also has 6 nm thick IGZO channel and 9 nm thick $Al_2O_3$ gate insulator, resulting in a $\lambda$ value of 8.5 nm. A series of comparisons are made between characteristics of 20-50 nm channel length nanospike FETs and conventional flat-edge FETs. The active semiconductor material used is indium gallium zinc oxide (IGZO); however, the designs proposed are equally applicable for other oxide semiconductors based FETs and indeed, single gate FETs with a variety of emerging semiconductors (29-32) and gate insulator materials (33). FETs with the new design can be conveniently combined in the same circuit with conventional FETs as the only changes are in electrode geometry, with all other layers being the same.

## Results and Discussion

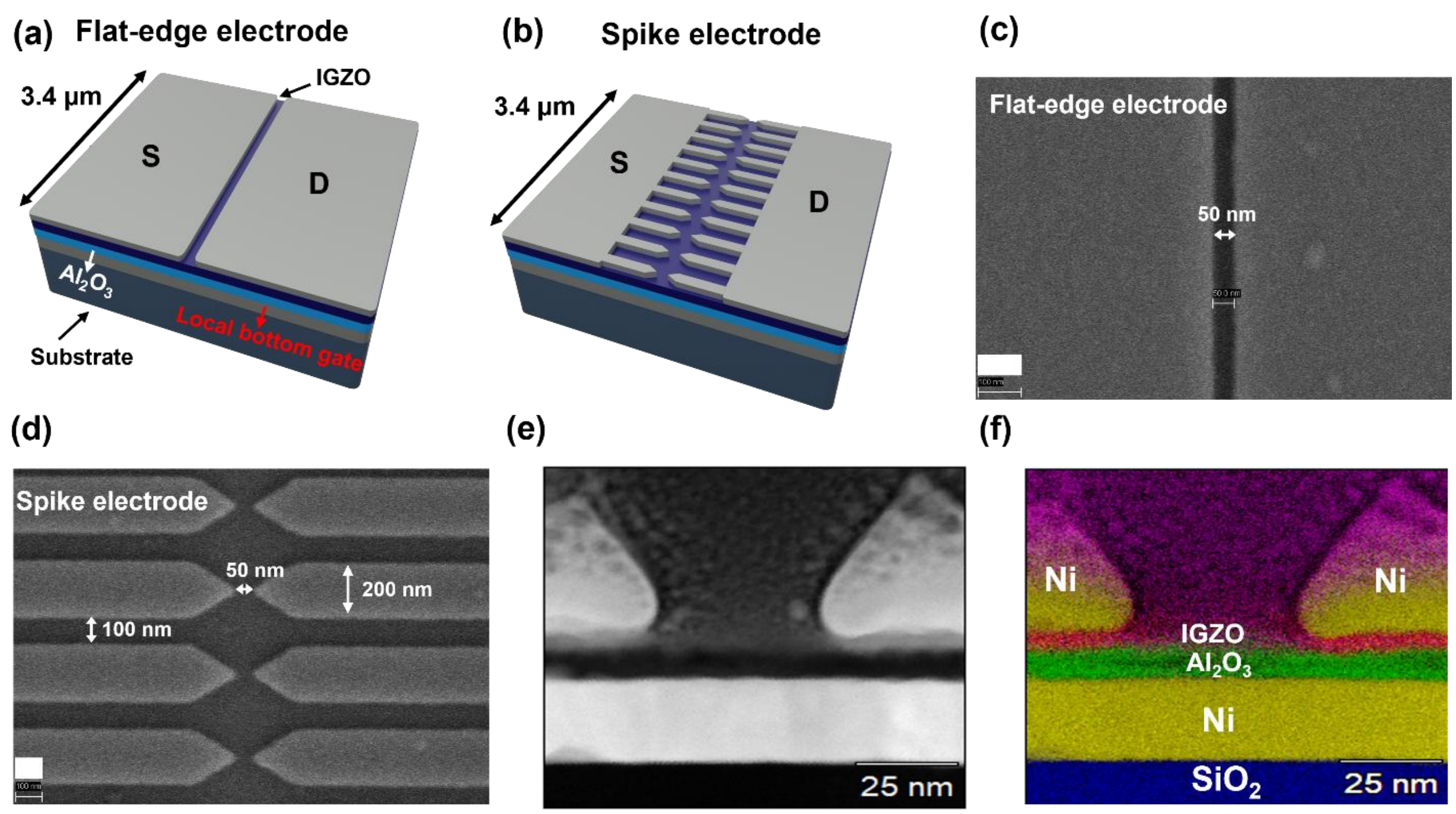

**Figure 1.** Schematic structures of bottom-gate, top-contact a-IGZO FETs with (a) conventional flat-edge and (b) nanospike shaped source/drain electrodes. Top-view scanning electron microscopy (SEM) images of 50 nm $L_{ch}$ devices with (c) flat-edge and (d) nanospike electrodes. (scale bar: 100 nm) (e) Cross-sectional Transmission Electron Microscopy (TEM) image with (f) energy dispersive X-ray spectroscopy (EDS) elemental mapping of the 50 nm $L_{ch}$ flat-edge device.

**Figure 1a** and **1b** illustrate schematic structures of the a-IGZO FETs with conventional flat-edge and nanospike-shape electrodes, respectively. The nanospike configuration proposed in this work consists of an array of nanospike-shaped electrodes with gaps in between, as illustrated in **Fig. 1b**. Detailed device fabrication steps are provided in the Methods section. As shown in **Fig. 1d**, each nanospike electrode consists of 200 nm wide metal electrode separated by 100 nm gap between adjacent electrodes. $L_{ch}$ of nanospike device is defined from tip to tip of nanospike electrode. Top-view SEM images illustrating overall $L_{ch}$ and $W_{ch}$ of both devices are shown in **Fig.S1**. The impact of nanospike electrode-to-nanospike electrode spacing on device performance at both 1000 nm and 50 nm $L_{ch}$ are further illustrated in **Fig. S2**. **Figure. 1e and 1f** show cross-sectional TEM images of representative 50 nm $L_{ch}$ flat-edge device with 9 nm thick $Al_2O_3$ and 6 nm thick IGZO. The EDS mapping images of each element are shown in **Fig. S3**. There is good agreement between target dimensions and layer thicknesses and measured dimensions and layer thicknesses.

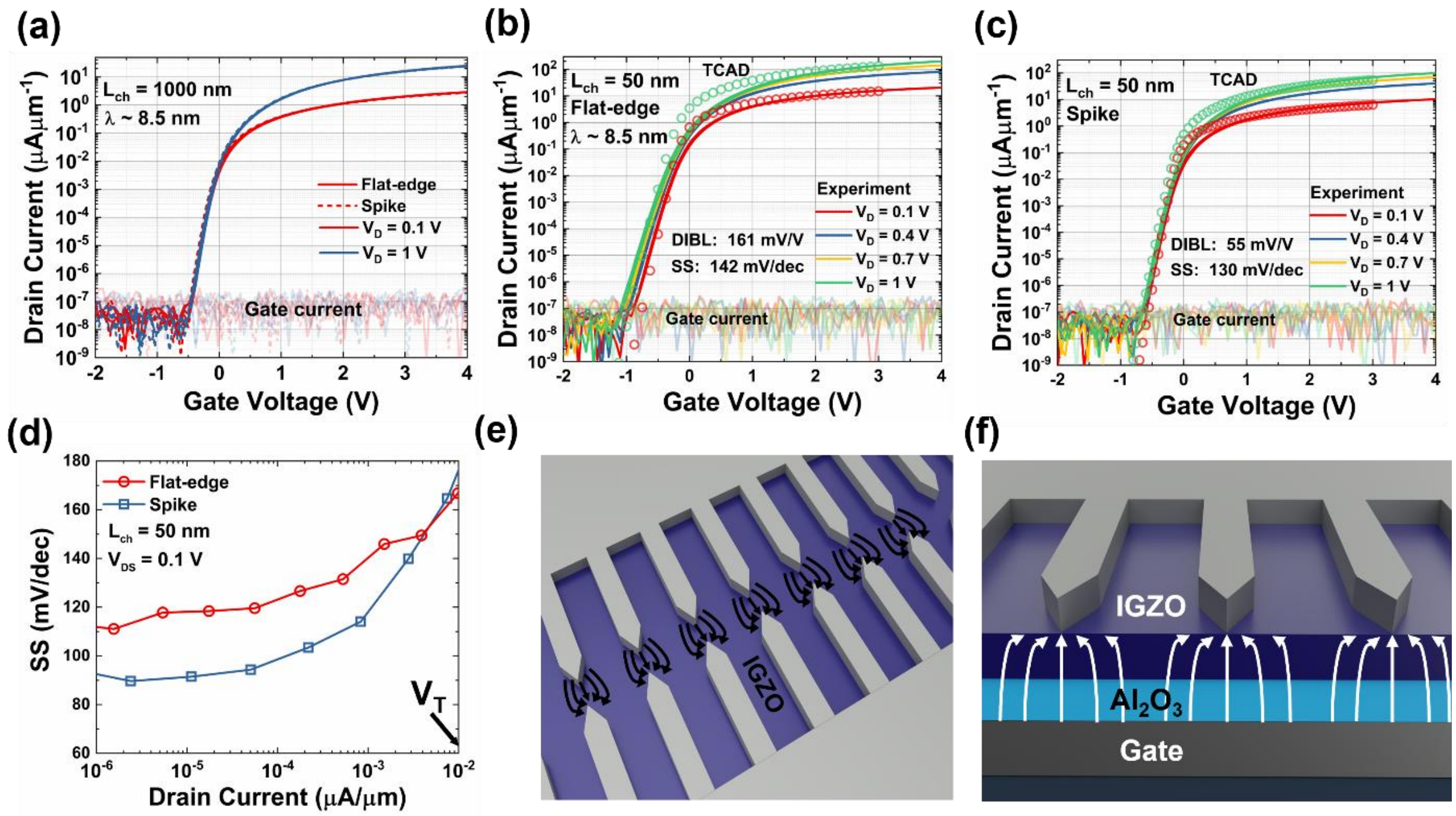

**Figure 2**. $I_D$-$V_G$ transfer characteristics of (a) nanospike and flat-edge device with $L_{ch}$ = 1000 nm and (b) a flat-edge device and (c) a nanospike device with $L_{ch}$ = 50 nm. SYNOPSYS SENTAURUS® TCAD simulated data at $V_{DS}$ = 0.1 V and 1 V are also illustrated in empty circle. The nanospike devices have significantly lower DIBL as well as lower SS compared to the conventional flat-edge electrode devices with same $L_{ch}$ (d) SS vs $I_D$ of 50 nm $L_{ch}$ devices during forward sweep. Schematic illustration of (e) lateral and (f) vertical electric field distributions in the nanospike device.

A comparison between the characteristics of nanospike FETs and those of conventional flat-edge electrode FETs are illustrated in **Fig. 2** and **Fig. 3**. **Figure 2** illustrates the $I_{DS}$ vs $V_{GS}$ of devices with nanospike and flat-edge electrodes at $L_{ch}$ = 1000 nm and 50 nm, highlighting the effect of the nanospike structure in mitigating SCEs. All drain current values are normalized by the channel width (3.4 μm) for both flat-edge and nanospike devices. For nanospike FETs, the channel width includes the spaces between individual nanospikes. Electrical measurements are performed under both forward/reverse voltage sweep, and all FETs exhibit negligible hysteresis and small gate leakage currents. The threshold voltage, $V_T$ is defined as a gate voltage at which drain current reaches 10 nAμm$^{-1}$ and $\lambda$ is estimated to be ~8.5 nm, using a relative dielectric constant of ~10 for IGZO (34). This method, or a closely related method, of defining $V_T$ is often used by researchers for convenience (10,35). At 1000 nm $L_{ch}$ (i.e., $L_{ch}/\lambda$ ~117), both devices exhibit negligible DIBL, implying strong gate control (**Fig. 2a**). Also, the nanospike and flat-edge devices show comparable $I_{DS}$. This is illustrated in additional detail in **Fig. S4a**. When $L_{ch}$ is reduced to 50 nm ($L_{ch}/\lambda$ ~5.88), flat-edge electrode FETs exhibit significant SCE with increased DIBL of 161 mV/V and SS of 142 mV/dec (**Fig. 2b**). In contrast, the nanospike device maintains a low DIBL of 55 mV/V and SS of 130 mV/dec, indicating improved immunity to SCEs (**Fig. 2c**). Also shown in **Fig. 2b** and **Fig. 2c** are the simulated transfer characteristics obtained with SYNOPSYS SENTAURUS®. Details of the simulation parameters are described in the Method sections. It can be seen that there is good agreement between the experimental and simulation data for both the flat-edge FET in **Fig. 2b** and the nanospike FET in **Fig. 2c**. **Figure 2d** compares SS vs $I_{DS}$, demonstrating the effectiveness of nanospike geometry in reducing SS compared to flat-edge electrode FETs.

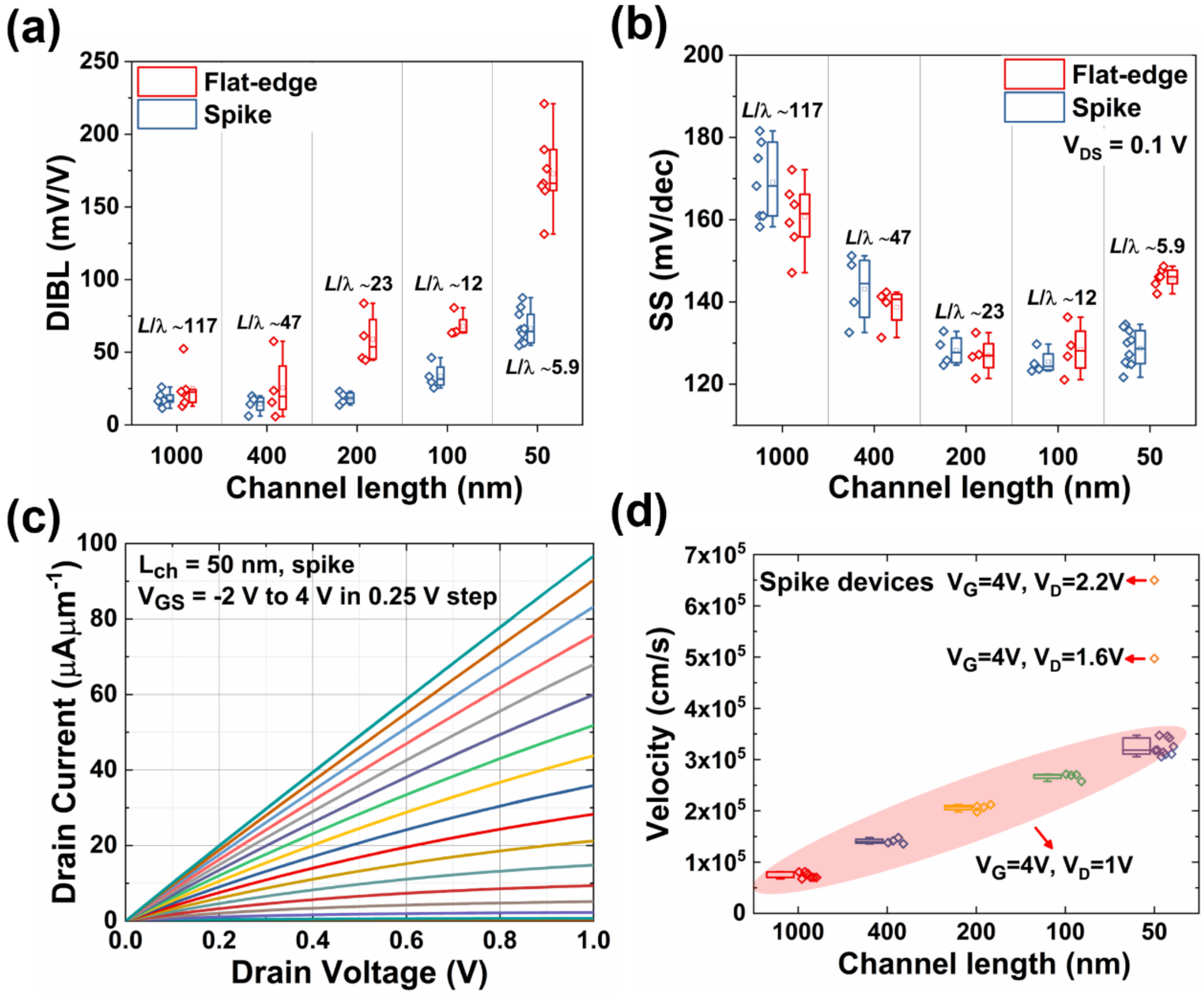

**Figure 3**. (a) DIBL vs $L_{ch}$,; also indicated are the corresponding $\lambda$ values for flat-edge FETs. The substantially lower DIBL values for the nanospike devices compared to the conventional flat-edge devices for the same channel length is clearly seen. (b) SS vs $L_{ch}$ of flat-edge and nanospike devices. The $\lambda$ values are for flat-edge FETs. (c) $I_D$-$V_D$ output characteristic of 50 nm $L_{ch}$ nanospike device. (d) Carrier velocity near the source side in nanospike devices in the linear region of operation for various channel lengths and bias conditions.

A second effect of the nanospike shape is enhancing gate control of the channel current. **Figure 2e and 2f** schematically illustrate the lateral and vertical electric field distribution in nanospike devices. This will also be illustrated in simulation data in **Figs. 5** and **Figs. 6**. The nanospike geometry can provide several advantages besides the ones described above. In addition to providing enhanced gate control over the channel current that effectively suppress SCEs at small $L_{ch}$, the nanospike electrode can reduce gate-to-source and gate-to-drain overlap areas. This reduction in overlap could lower parasitic overlap capacitance, thereby enhancing the switching speed of device. The trends in DIBL and SS as a function of channel length are illustrated in **Fig. 3a** and **3b** for nanospike and flat-edge FETs. DIBL is comparable for nanospike and flat-edge FET at larger channel lengths above 200 nm, whereas at $L_{ch}$ = 200 nm and lower, DIBL values are substantially lower for nanospike geometry FETs. The SS values are comparable at channel lengths above 100 nm, and diverge significantly at 50 nm channel lengths, with the nanospike FETs maintaining a low SS value, while the flat-edge FETs have markedly higher SS values. $V_T$ values of both devices at various $L_{ch}$ are illustrated in **Fig. S5**. The output characteristics ($I_{DS}$ vs $V_{DS}$) of a 50 nm $L_{ch}$ nanospike device is shown in **Fig. 3c**, exhibiting maximum $I_{DS}$ of 96.7 μAμm$^{-1}$ under $V_{DS}$ = 1V and $V_{GS} - V_T = 4.17$ V. In **Fig. 3d,** the carrier velocity near the source side extracted from linear-regime ($v = I_{DS}/(WC_{ox}(V_{GS} - V_T))$) is plotted as a function of $L_{ch}$. A peak carrier velocity of $6.5 \times 10^5$ cm/s is achieved under $V_G = 4$V and $V_D = 2.2$ V with superior gate control even at large drain bias, as shown in **Fig. S6**. In calculating carrier velocity, contact resistance was neglected.

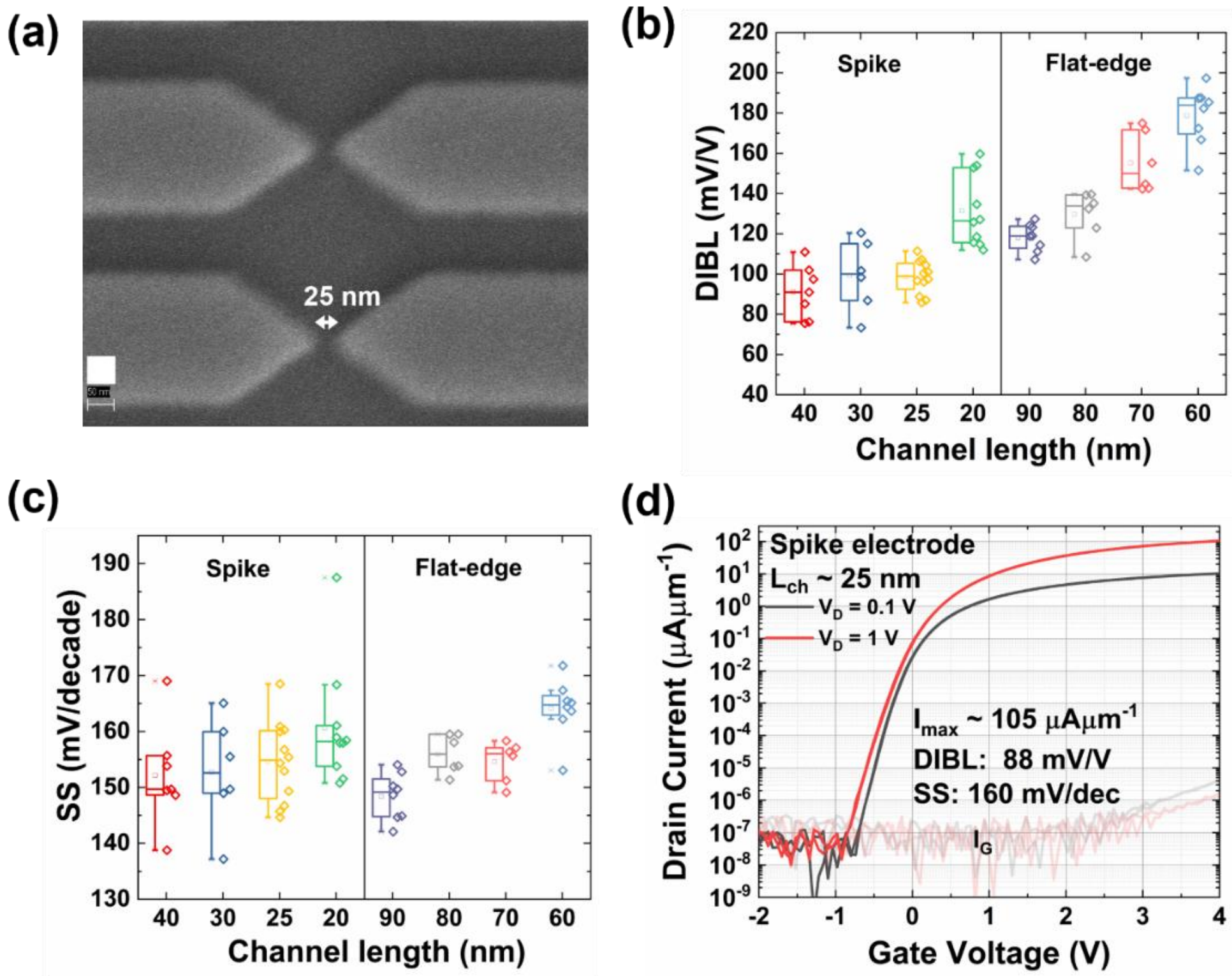

**Figure 4**. (a) Top view SEM image of a 25 nm $L_{ch}$ nanospike FET. (scale bar: 50nm) (b) DIBL vs $L_{ch}$, (c) SS vs $L_{ch}$ of flat-edge and nanospike FET. (d) $I_D$-$V_G$ transfer characteristic of 25 nm $L_{ch}$ nanospike FET.

Performance of nanospike devices at channel lengths in the range 20—40 nm is illustrated in **Fig. 4**. All devices reported in **Fig. 4** were fabricated in the same batch. For comparison, flat-edge device characteristics are shown for channel lengths in the range 60-90 nm, which provide an appropriate basis to compare with nanospike FETs. The yield of flat-edge FETs with channel lengths below 50 nm was very small due to fabrication constraints. For this reason, data for flat-edge FETs with channel lengths in the range 20-40 nm could not be shown in **Fig. 4**. It is seen in **Fig. 4b** that the DIBL value of 20 nm channel length nanospike FETs is roughly comparable to that in 70–80 nm channel length flat edge FETs, providing a remarkable illustration of the utility and efficacy of the nanospike geometry FETs proposed in this work. The SS value for the 20 nm nanospike FET, shown in **Fig. 4c,** is comparable to that of much larger channel length flat-edge electrode FETs, again demonstrating the advantages of the nanospike FET vis-à-vis the conventional design. Importantly, these advantages become ever more pronounced as the channel lengths are reduced to result in $L_{ch}/\lambda$ values of $< 5$. In this regime, conventional single gate FETs typically possess inadequate gate control. This is evident, for example, in recent publications on single gate IGZO FETs with channel lengths below 50 nm. (18) For this technologically crucial range of channel lengths, nanospike electrode FETs will permit the use of single gate geometries without compromising the device characteristics. The comparable magnitudes of SCE in 20 nm channel length nanospike FETs and ~70 nm flat-edge FETs underline the utility of the nanospike geometry in practical applications. The transfer and output characteristics of a 25 nm $L_{ch}$ nanospike device are shown in **Fig. 4d** and also in **Fig. S7**. The device has on-current of ~105 μAμm$^{-1}$ with a low DIBL (88 mV/V) and SS (160 mV/dec), confirming that the nanospike structure supports aggressive $L_{ch}$ scaling without compromising key electrical metrics. A benchmark table comparing nanospike device with other sub-50 nm $L_{ch}$ oxide semiconductor devices is provided in supplementary **Table S1**.

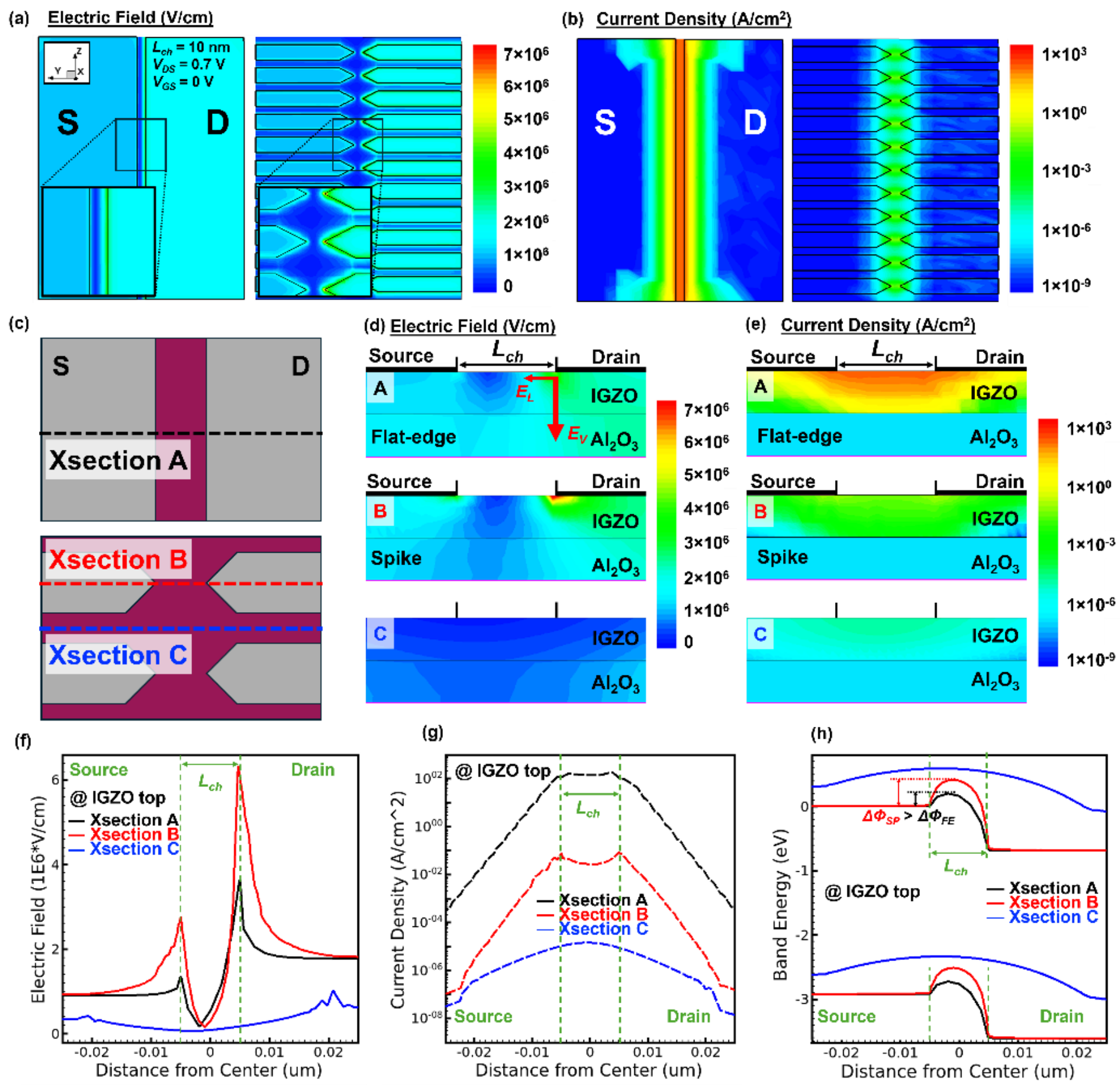

**Figure 5.** (a) Electric field and (b) current density distributions of both devices in the subthreshold regime, observed at the top surface of IGZO channel. (c) Schematic of cut-lines along the channel direction for flat-edge and nanospike device. For the nanospike device, cross-section B corresponds to the nanospike tip and cross-section C corresponds to the gap between two nanospike electrodes. Two-dimensional (d) electric field and (e) current density distributions for each cross-section. (f) Electric field and (g) current density profiles along cut-lines A, B and C at the top surface of IGZO channel (as defined in **Fig. 5c**). (h) Band energy profiles of both devices measured at top of the IGZO channel. The energy barriers for the flat-edge and nanospike device are expressed as $\Delta\Phi_{FE}$ and $\Delta\Phi_{SP}$, respectively.

To gain deeper insight into electron transport and electric field distributions of both devices, three-dimensional simulations were performed using Synopsys Sentaurus Technology Computer-Aided Design (TCAD) (**Figs. S8**, **Figs. S9**, **Figs. 5 and Figs. 6**). First, simulation results for 50 nm $L_{ch}$ nanospike and flat-edge devices are performed, as shown in **Fig. 2a** and **Fig. 2b** alongside experimental data. Key simulation parameters were selected to as to result in good agreement between experimental and simulated transfer characteristics. A complete set of simulation parameters are listed in the Methods section. To predict the characteristics of highly scaled devices–both flat edge and nanospike with channel lengths of 10-20 nm, which are less than what are currently achievable with our fabrication process, are also simulated. These FETs possess $L_{ch}/\lambda$ values of 5, 3.75, 2.5. It is observed that nanospike devices continue to exhibit improved short-channel-effects even at L = 10 nm corresponding to $L/\lambda = 2.5$. (**Figs. S9**). **Figure 5** shows electric field distributions, current densities, and band-edge potential profiles of both devices in the subthreshold-regime. In the subthreshold regime, for bottom gate-top contact FETs, leakage currents can flow in the top part of the IGZO channel (i.e., away from the bottom gate and gate insulator) and needs to be minimized. **Figure 5a** and **Figure 5b** show the top-view electric field and current density distributions of both devices at the top surface of IGZO channel. In the flat-edge device, the lateral electric field is uniformly high across the channel. In contrast, the nanospike device exhibits position-dependent electric field distribution due to its electrode geometry. The drain-side tips exhibit enhanced electric field compared to conventional flat-edge device. Owing to this electric field difference along the top surface of the IGZO channel, the current distribution difference of both devices is clearly observed in **Fig. 5b**. The flat-edge device has a higher current between source and drain electrodes, whereas the nanospike device has lower overall current, with the current primarily confined between the nanospike tips. Thus, the sub-threshold swing is expected to be steeper for nanospike FETs as the effective area of conduction is smaller and confined to the region between the nanospike tips along the source-drain direction, which may reduce the number of interface traps that influence device operation. **Fig 5c** illustrates three specific cut lines (A, B and C) in the devices. **Figures 5d** and **5e** show the cross-sectional contour plots of the electric field and current density respectively, along the three cut-lines, A, B, and C defined in **Fig. 5c**. **Figure 5e** clearly shows that there are substantially larger currents flowing at the top surface of the channel for the flat-edge FET(cut line A) compared to the nanospike FET (cut line B). This is the main reason for the superior sub-threshold current control and smaller DIBL in nanospike FETs compared to flat-edge FETs. This is more clearly seen in **Fig. 5g** in which the current density is plotted as along lines A,B, and C. It is noteworthy that the field at the source edge and drain edge are higher and the field near the center of the channel is lower for the nanospike FET compared to the flat-edge FET in the sub-threshold region (**Fig.5d** and **5f**). This difference in electric field distribution affects energy barrier height, as shown in **Fig. 5h.** Specifically, flat-edge device has a lower energy barrier height compared to the nanospike device, which in turn allows more leakage current to flow. In the on-state, this is no longer the case, as illustrated in **Fig. 6.**

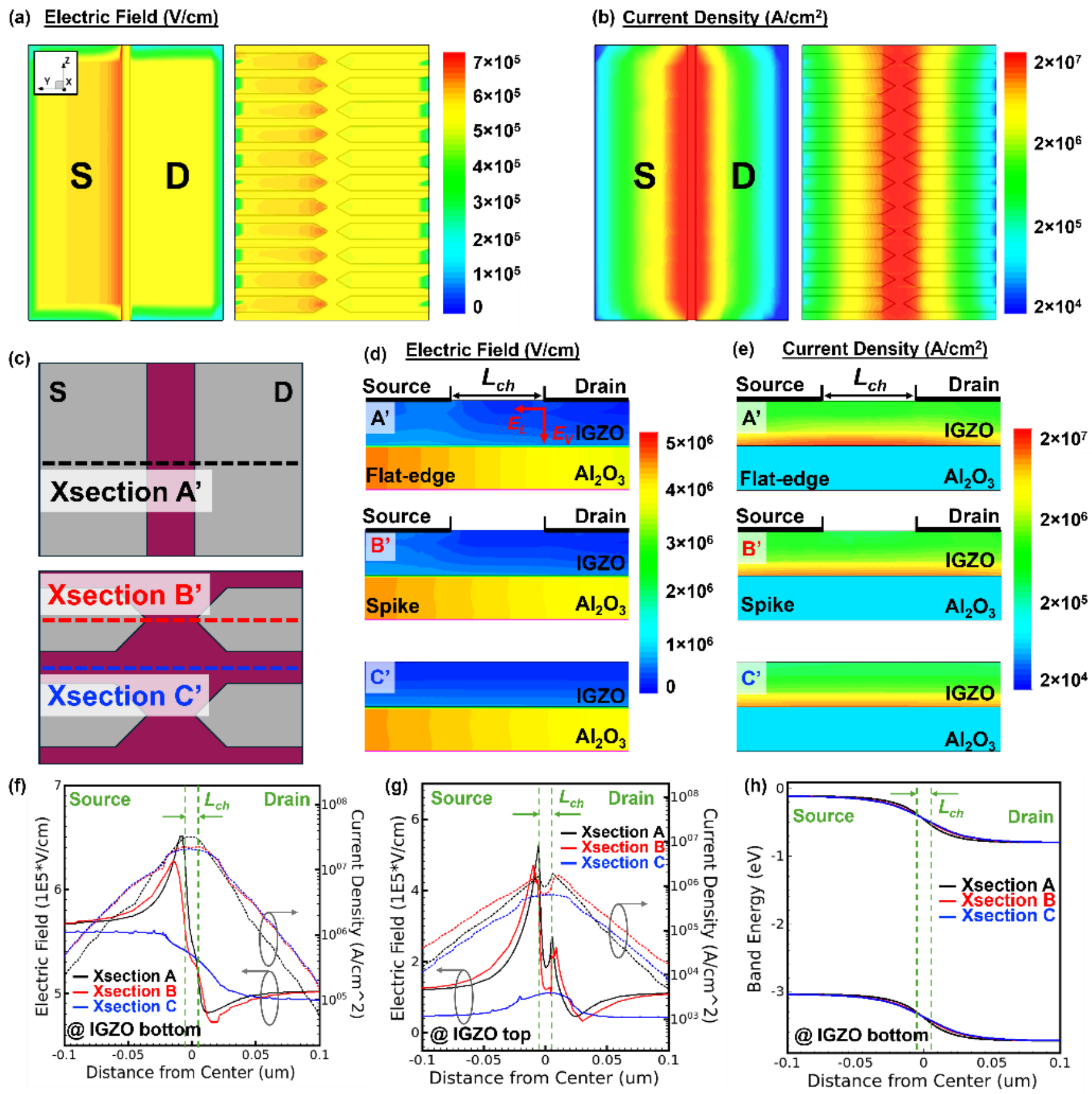

**Figure 6.** (a) Electric field and (b) current density distributions in the on-state ($V_{GS}$ = 3 V, $V_{DS}$ = 0.7 V), observed at the interface between the bottom of IGZO channel and insulator/gate electrode. The flat-edge and nanospike electrodes are overlaid as reference to indicate the channel region. (c) For the nanospike device, cross-section B corresponds to the nanospike tip and cross-section C corresponds to the gap between two nanospike electrodes. (d) Electric field and (e) current density distributions for each cross-section. Electric field and current density profiles along the cut-lines A', B', and C' (as defined in **Fig. 6c**) (f) at the bottom of IGZO channel and (g) at the top of IGZO channel. (h) Band energy profiles of both devices measured at the bottom of IGZO channel..

On the other hand, in the on-state of the devices, when both $V_{GS}$ and $V_{DS}$ are high, the main current path is located at the interface between the bottom of the IGZO channel and insulator/gate electrode. The electrical properties of both devices in the on-state are examined, as shown in **Figs. 6**. The flat-edge and nanospike electrodes are overlaid as reference to indicate the channel region. In contrast to the subthreshold regime–where electric fields are higher for the nanospike FET near the electrode edges but smaller within the channel, the electric fields are higher for the flat-edge

FETs throughout the channel especially at the bottom of IGZO channel (**Figs. 6a~6g**). **Fig. 6b** indicates that the transfer length in both nanospike and flat-edge FETs are comparable and are roughly in the range of 25-50 nm. Quantitative comparisons in **Fig. 6f** show that the electric field and current density of the flat-edge device are only slightly higher than those of the nanospike device at the bottom of IGZO channel. In flat-edge FETs, in the on state, the electric field near the drain is much higher near the top surface of the channel (**Fig. 6g**) This leads to enhanced charge transport in the channel away from gate insulator interface and hence less efficient gate control. In comparison, for nanospike FETs, the electric field is much lower near the drain and towards the top of the channel resulting in comparatively less conduction in these regions. This is one of the reasons why on-currents are slightly lower for nanospike FETs compared to flat-edge FETs. The band energy profiles are shown in **Fig. 6h.**

It can be observed in **Fig. 6f** and also deduced from a comparison of **Fig. 2b** and **2c**, that the nanospike device exhibits slightly lower $I_{DS}$ compared to the flat-edge electrode FET. This is a result of a combination of several factors including: i) a noticeable negative $V_T$ shift in the flat edge FET due to severe DIBL; ii) a possibly smaller effective channel width in the nanospike device iii) a reduced effective carrier velocity in the nanospike structure due to lower average electric fields; and (iv) higher DIBL in the flat-edge electrode FETs resulting in higher currents due to injection and transport in the bulk of the semiconductor (which is 6 nm thick), away from the interface.

To understand more deeply which design features of the nanospike electrode FET are responsible for this excellent performance at smaller channel lengths, comparisons were made of characteristics of two other geometries related to that of the nanospike FETs reported in this paper. These diagnostic geometries are illustrated in **Fig. 7a** and **7b.** All devices reported in **Fig. 7** were fabricated in the same batch, as described in the methods section below, but without encapsulation with PMMA. These diagnostic FET geometries are designated as finger electrode FETs and short nanospike FETs, as illustrated in **Fig. 7a** and **7b**. The finger electrode FET has the same ratio of electrode spacing to electrode width that the nanospike FETs reported above possess. The only difference is that the electrode tips are flat-edged instead of tapered. The short nanospike geometry is similar to the previously described nanospike FETs except that the nanospike region is less elongated and more compact and hence takes up less area, which is important for increasing the density of FETs that can be fabricated in a given area. As summarized in **Figs. 7c** and **7d**, both long and short nanospike configurations exhibit lower DIBL and SS compared to flat-edge and finger electrode devices at $L_{ch}$ = 50 nm. Interestingly, the short nanospike devices show suppressed SCEs which are comparable to that in long-nanospike devices. In contrast, the finger-shaped electrodes exhibit pronounced SCEs, with DIBL and SS values similar to those of flat-edge electrode devices. (**Fig. S10)** This clearly indicates that the spacing between electrodes, which finger electrode FETs have in common with nanospike FETs, is in itself insufficient to suppress SCEs, thereby highlighting the importance of electrostatic control provided by the tapered end of the nanospike geometry devices (in both short nanospike and long nanospike FETs). The similarity between the characteristics of the short nanospike and long nanospike FETs show that the tapering is the crucial aspect of the design that provides better gate control and reduced SCE. This tapered geometry is crucial for better gate control of drain current and especially at small channel lengths. TCAD simulation results of scaled flat-edge, nanospike, and finger electrode devices are illustrated in **Fig. S11** and are consistent with the experimental data. The data shows that with the use of nanospike electrodes, the simpler single-gate geometry can be used down to smaller channel lengths than is possible with conventional FET geometries, obviating the need for multiple gate and three-dimensional geometries. This advantage can be very useful in designing dense, multi-level, BEOL FET circuitry for future applications.

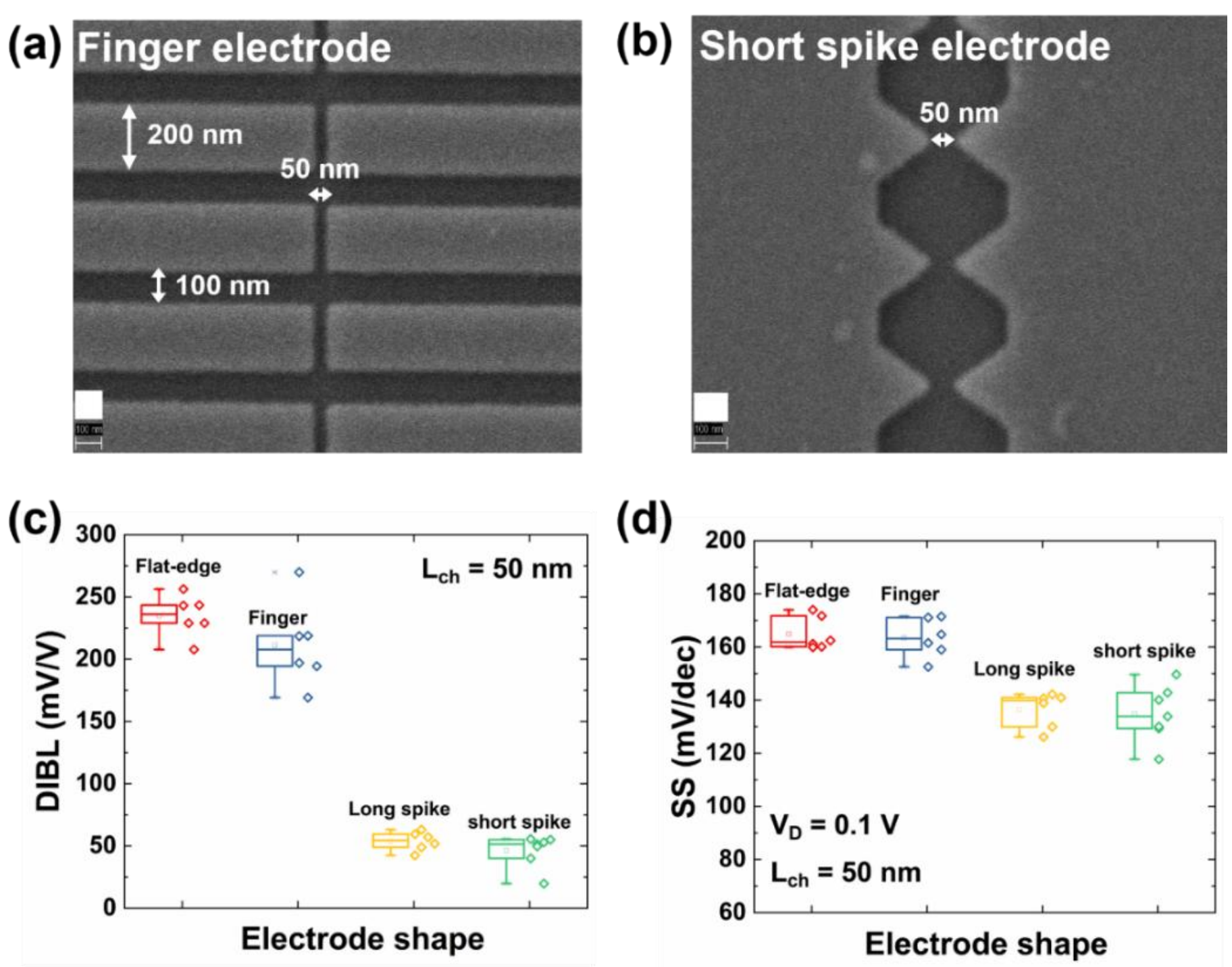

**Figure 7**. Top-view scanning electron microscopy (SEM) images of 50 nm $L_{ch}$ devices with (a) finger electrode and (b) short nanospike electrode. (scale bar: 100 nm). (c) DIBL and (d) SS comparisons for various electrode geometries at $L_{ch}$ of 50 nm. The SEM image of the long nanospike electrode device is shown in **Fig. 1**(d)

## Conclusion

This study demonstrates that nanospike electrodes can reduce short channel effects in a-IGZO FETs compared to conventional flat-edge electrodes. By comparing DIBL and sub-threshold swing across various $L_{ch}$ and electrode geometries, it has been established that the nanospike electrode geometry results in an effective reduction in characteristic length ($\lambda$), which is highly beneficial for scaling down the $L_{ch}$ of transistors without compromising performance. The DIBL in a 20-25 nm channel length nanospike electrode FETs is comparable to that in a much larger channel length (70-80 nm) conventional flat-edge electrode FETs that are fabricated alongside. These findings highlight the nanospike geometry as a promising device design for device scaling in BEOL compatible amorphous oxide FETs. Moreover, the nanospike electrode design concept is likely to also benefit a wide range of emerging semiconductor-based short $L_{ch}$ FETs and can be extended to multi-gate topologies as well. Simulations with SYNOPSYS SENTAURUS® provided insight into the physics of operation of these devices and confirmed that the improvements in nanospike FETs are related to the better gate control of channel current. Importantly, the simulations indicate that the advantages of nanospike electrode FETs will not only persist but become more pronounced when channel lengths are reduced below 25 nm down to 10 nm.

## Methods

### Device Fabrication

First, an 86 nm $SiO_2$ layer is thermally grown on $p^{++}$ Si substrates. The 25 nm nickel (Ni) local bottom gate electrode is defined using electron beam lithography (JBX-8100FS/E) and electron beam evaporation system (PRO Line PVD 200). Following this, a 9 nm $Al_2O_3$ gate insulator layer is deposited via atomic layer deposition at 200 ºC (Fiji F200). A 6 nm thick IGZO channel is deposited above the gate insulator by RF-sputtering at 150 W at 5 mTorr with 7% $O_2$ pressure (PVD75). The IGZO deposition is done at room temperature with a base pressure of $\sim 2.5 \times 10^{-3}$ Torr. After deposition, the devices are annealed on a hotplate at 350 ºC for 1 hour in air. Then, the IGZO channels are patterned using electron beam lithography and wet-etching using a diluted HCl etch solution. Source and drain electrodes (Ni) are patterned using the electron beam lithography and lift-off. For source/drain formation, diluted ZEP (ZEP520A:anisole = 1:2) resist is spun at 2000 rpm for 60 s and baked at 180 °C for 2 minutes. An electron beam energy of 100 keV is used for lithography. ~30 nm Ni is deposited via electron beam evaporation system, followed by lift-off in Remover PG. Finally, PMMA A4, which serves as an encapsulation layer, is spin-coated and annealed at 180 ºC for 2 minutes. Then, windows in PMMA are defined for contacting gate, source and drain electrodes. All fabrication processes are performed at temperature below 350 ºC, making the entire process compatible with BEOL requirements.

**Device Characterization**

Electrical measurements are performed with an Agilent 4155C semiconductor parameter analyzer, at room temperature, in a dark environment and under vacuum (~$6\times10^{-5}$ Torr).

**Scanning electron microscopy**

SEM images of the IGZO FETs used in this study are obtained using a Zeiss Gemini 460 SEM system at an accelerating voltage of 5 kV.

**3D TCAD simulation**

Synopsys Sentaurus 3D TCAD simulations are used to simulate 50 nm $L_{ch}$, 3.4 µm $W_{ch}$ devices using following parameters: IGZO dielectric constant: 15, $Al_2O_3$ dielectric constant: 10, IGZO channel donor concentration: $4\times10^{18}$ $cm^{-3}$, IGZO bandgap: 3.0 eV, IGZO electron affinity: 4.1 eV, IGZO electron mobility: 5 $cm^2V^{-1}s^{-1}$, gate work function: 5 eV and source/drain contact resistivity: $1\times10^{-7}$ $\Omega cm^2$. Scaled devices (10 nm $L_{ch}$, 0.34 µm $W_{ch}$, 4 nm $T_{ox}$, 4 nm $T_{IGZO}$) used following parameters: IGZO dielectric constant: 10, $Al_2O_3$ dielectric constant: 10, IGZO channel donor concentration: $4\times10^{18}$ $cm^{-3}$, IGZO bandgap: 3.0 eV, IGZO electron affinity: 4.1 eV, IGZO electron mobility: 5 $cm^2V^{-1}s^{-1}$, gate work function: 5 eV and source/drain contact resistivity: $1\times10^{-7}$ $\Omega cm^2$. Simulated curves with different mobility and contact resistivity are illustrated in **Fig. S12**

**Supporting Information**

Electrical properties of nanospike electrode devices with various spacing between nanospike electrodes, cross-sectional TEM images with EDS elemental mapping, field effect mobility of device, electrical properties of $Al_2O_3$ capacitor, $V_T$ of devices at various $L_{ch}$, $I_D$-$V_G$ curve of 50 nm $L_{ch}$ nanospike device at $V_D$ = 0.1 V and 2.2 V, $I_D$-$V_D$ curve of 25 nm $L_{ch}$ nanospike device, $I_D$-$V_G$ of 50 nm $L_{ch}$ finger and short-nanospike device. TCAD simulation results

**AUTHOR INFORMATION**

**Corresponding Author**

Ananth Dodabalapur − Chandra Family Department of Electrical and Computer Engineering, The University of Texas at Austin, Austin, Texas 78712, USA; Microelectronics Research Center, The University of Texas at Austin, Austin, Texas 78758, United States; Email: ananth.dodabalapur@engr.utexas.edu

**Authors**

Chankeun Yoon − Chandra Family Department of Electrical and Computer Engineering, The University of Texas at Austin, Austin, Texas 78712, USA; Microelectronics Research Center, The University of Texas at Austin, Austin, Texas 78758, USA

Juhan Ahn − Chandra Family Department of Electrical and Computer Engineering, The University of Texas at Austin, Austin, Texas 78712, USA

Yuchen Zhou − Chandra Family Department of Electrical and Computer Engineering, The University of Texas at Austin, Austin, Texas 78712, USA; Microelectronics Research Center, The University of Texas at Austin, Austin, Texas 78758, USA

Jaydeep P. Kulkarni − Chandra Family Department of Electrical and Computer Engineering, The University of Texas at Austin, Austin, Texas 78712, USA

**Author Contributions**

† C.Y and J.A equally contributed to this work. C.Y., J. A. and A.D. conceived the experiments and wrote the manuscript with input from authors. C.Y. fabricated devices and performed electrical characterization. J.A. and J.P.K performed simulation. Y. Z. assisted with IGZO deposition. The manuscript was written through contributions of all authors. All authors have given approval to the final version of the manuscript.

Data availability

All data are available in the paper and Supplementary Information are available from the corresponding author upon reasonable request.

**ACKNOWLEDGMENT**

The authors would like to thank Dr. Ioana R. Gearba-Dolocan and Dr. Xun Zhan at University of Texas at Austin (Texas Materials Institute) for TEM sample preparation and characterization. This work was supported by the National Science Foundation under Grant No. CMMI-1938179, by Keck foundation under grant #26753419 and by National Science Foundation cooperative agreement EEC-1160494 Semiconductor Research Corporation (SRC) task ID no. 2962.001. The work was done at the Texas Nanofabrication Facility supported by the National Science Foundation under Grant NNCI-2025227.

Supporting information for

# Enhancing gate control and mitigating short channel effects in 20-50 nanometer channel length amorphous oxide Thin-Film Transistors

Chankeun Yoon[1,2,†] Juhan Ahn,[1,†] Yuchen Zhou,[1,2] Jaydeep P. Kulkarni,[1] and Ananth Dodabalapur[1,2,*]

[1] *Chandra Family Department of Electrical and Computer Engineering, The University of Texas at Austin, Austin, Texas 78712, USA*

[2]*Microelectronics Research Center, The University of Texas at Austin, Austin, Texas 78758, USA*

[†] *These authors equally contributed to this work*

***Email: ananth.dodabalapur@engr.utexas.edu***

**Supporting Information 1**

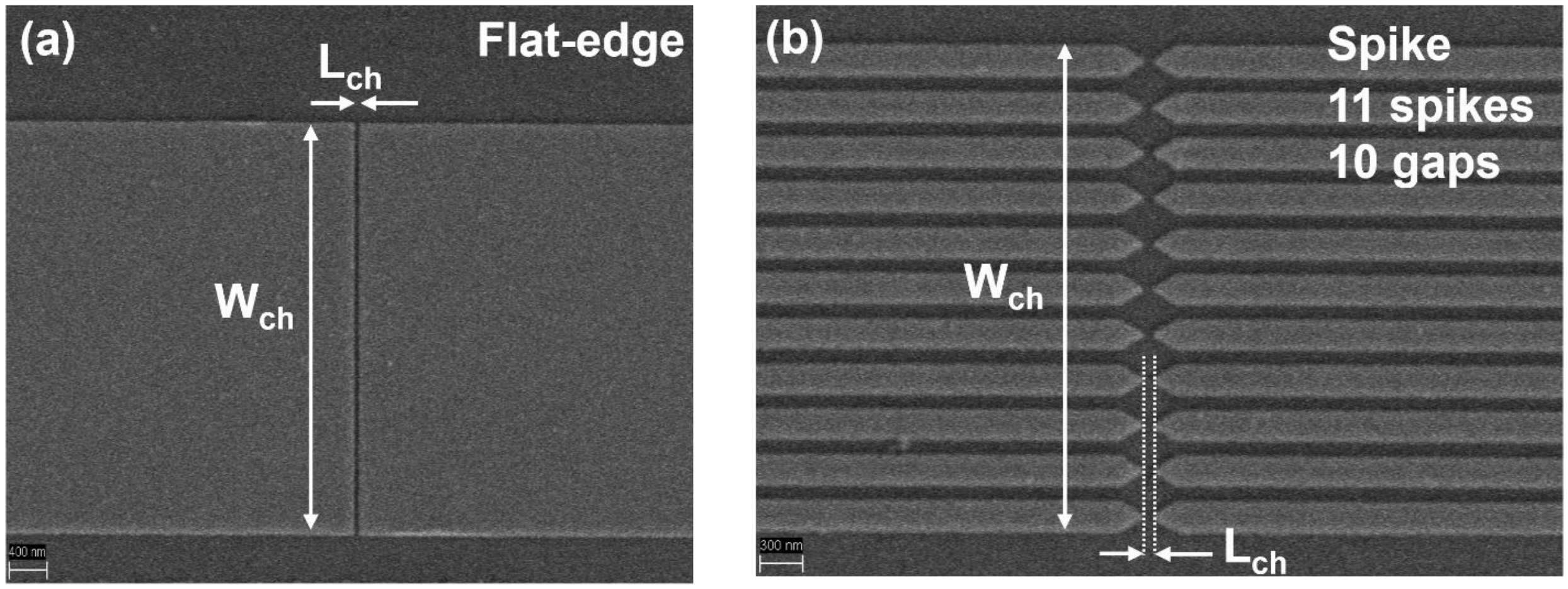

**Figure S1**. Top view SEM image of (a) flat-edge and (b) nanospike device.

**Supporting Information 2**

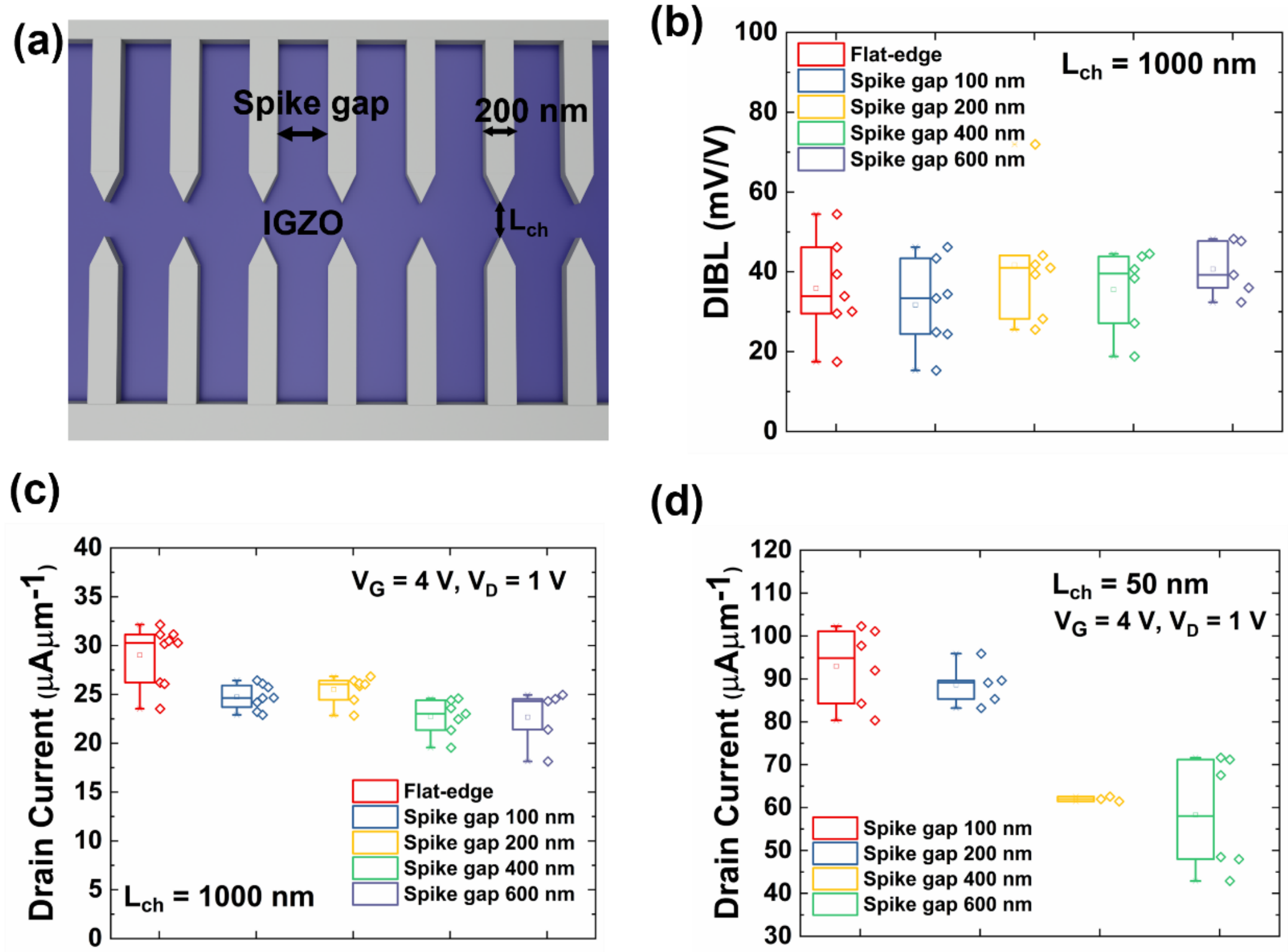

**Figure S2**. (a) Design of nanospike-shaped electrodes. (b) DIBL and (c) drain current comparison of 1000 nm $L_{ch}$ devices with flat-edge electrode and nanospike electrode with various gaps between adjacent nanospike electrodes. (d) Drain current comparison of 50 nm $L_{ch}$ nanospike devices with various gaps.

**Figure S2a** illustrates the schematic structure of the nanospike-shaped electrodes, where the gap between adjacent nanospike electrode is changed from 100 nm to 600 nm. In this comparison, the samples were not passivated with PMMA but were measured under vacuum conditions. **Figure S2b** and **S2c** shows the DIBL characteristics and $I_{DS}$ of 1000 nm $L_{ch}$ devices. Since $L_{ch}/\lambda$ is large, all devices exhibit small DIBL values (< 100 mV/V) and the impact of gap variation is not significant. However, as the gap decreases from 600 nm to 100 nm, the $I_{DS}$ increases, as shown in **Fig. S2c**. This trend becomes more pronounced at 50 nm $L_{ch}$, as shown in **Fig. S2d**. Based on these results, a 100 nm gap between nanospike electrodes is chosen for this study.

**Supporting Information 3**

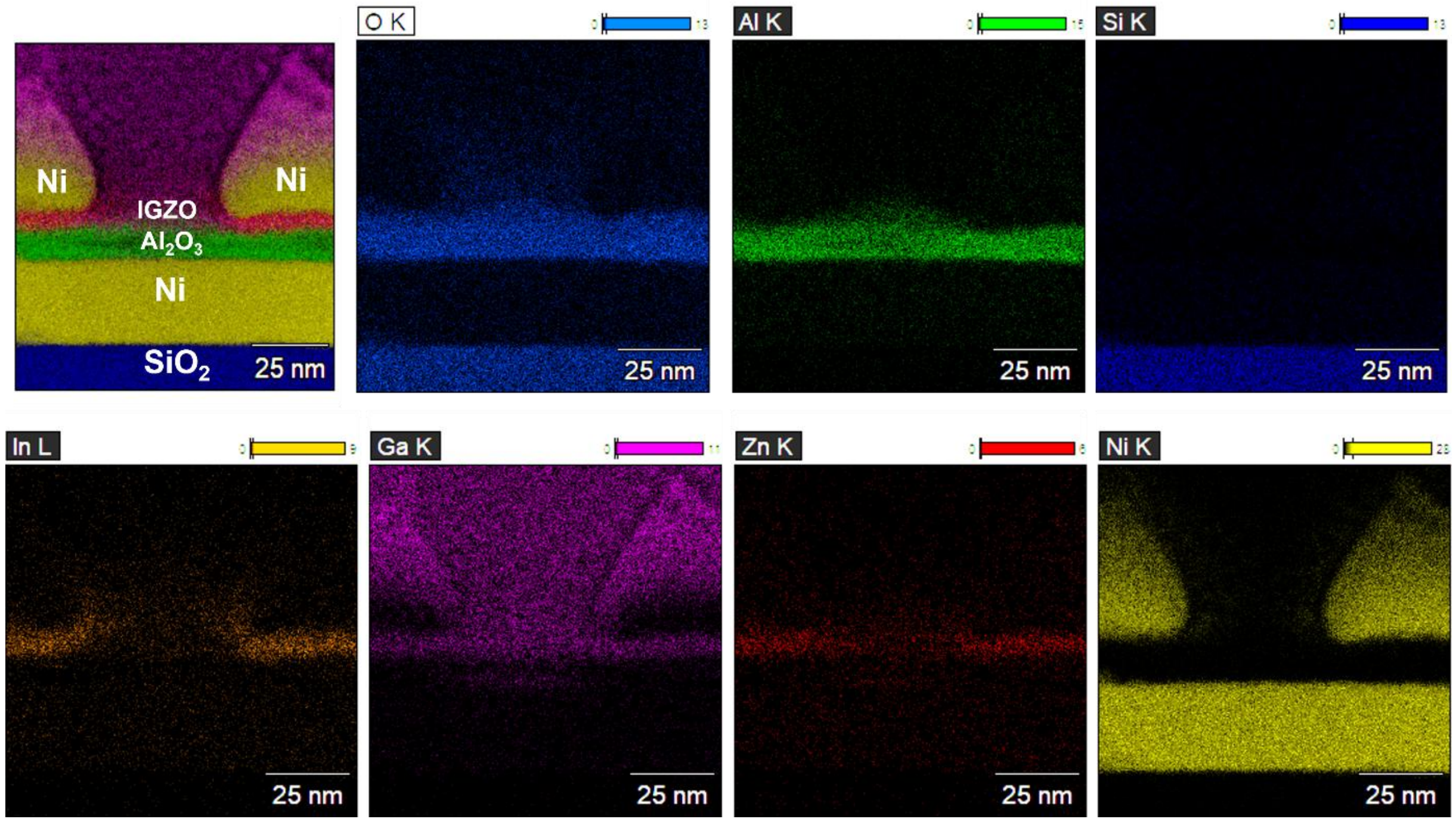

**Figure S3**. Cross-sectional Transmission Electron Microscopy (TEM) images with energy dispersive X-ray spectroscopy (EDS) elemental mapping of the 50 nm $L_{ch}$ flat-edge device. Ga distribution is influenced during FIB cutting which used Ga ion for TEM sample preparation.

**Supporting Information 4**

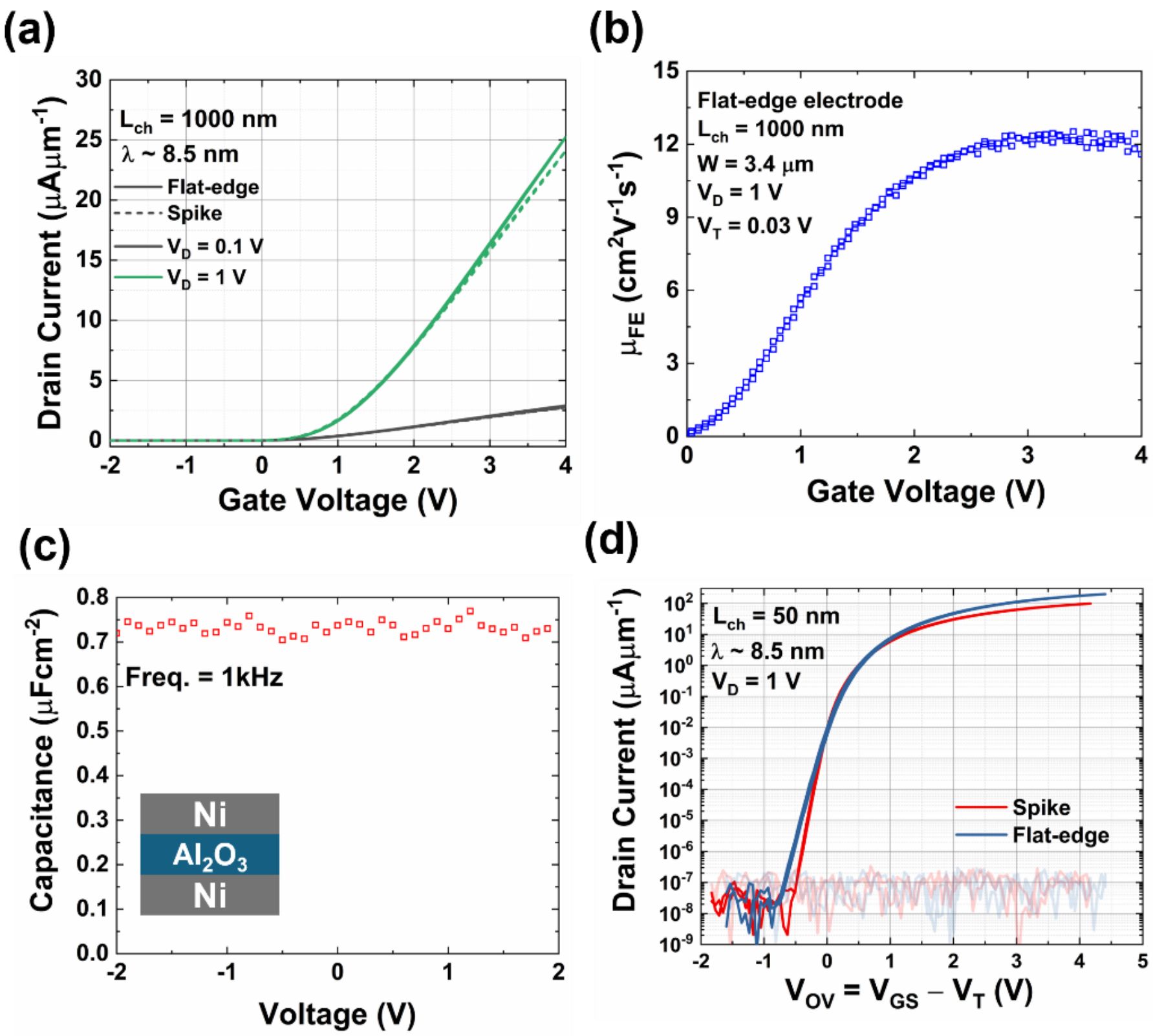

**Figure S4**. (a) $I_D$-$V_G$ transfer characteristics of 1000 nm $L_{ch}$ flat-edge and **nano**spike device, plotted in a linear scale. (b) Field effect mobility of 1000 nm $L_{ch}$ flat-edge device. The calculated average mobility is 12 cm$^2$V$^{-1}$s$^{-1}$ at $V_G$ = 4 V and $V_D$ = 1V. (c) Measured capacitance vs. voltage characteristics of Ni/$Al_2O_3$/Ni capacitor at 1 kHz. (d) $I_D$-$V_{ov}$ characteristics of 50 nm $L_{ch}$ flat-edge and nanospike device.

**Figure S4a** illustrates $I_D$-$V_G$ of 1000 nm $L_{ch}$ flat-edge and nanospike devices in linear scale showing comparable current. Field-effect mobility, $\boldsymbol{\mu_{FE}}$ of 1000 nm $L_{ch}$ flat-edge device is plotted in **Fig. S4b** ($\boldsymbol{\mu_{FE}}$ = $g_m L/(WC_{ox}V_D)$). **Fig. S4c** shows capacitance vs. voltage of 9 nm thick $Al_2O_3$ used in this study to calculate relative dielectric constant of $Al_2O_3$. Steeper turn-on behavior of 50 nm $L_{ch}$ nanospike device is illustrated in **Fig. S4d**.

**Supporting Information 5**

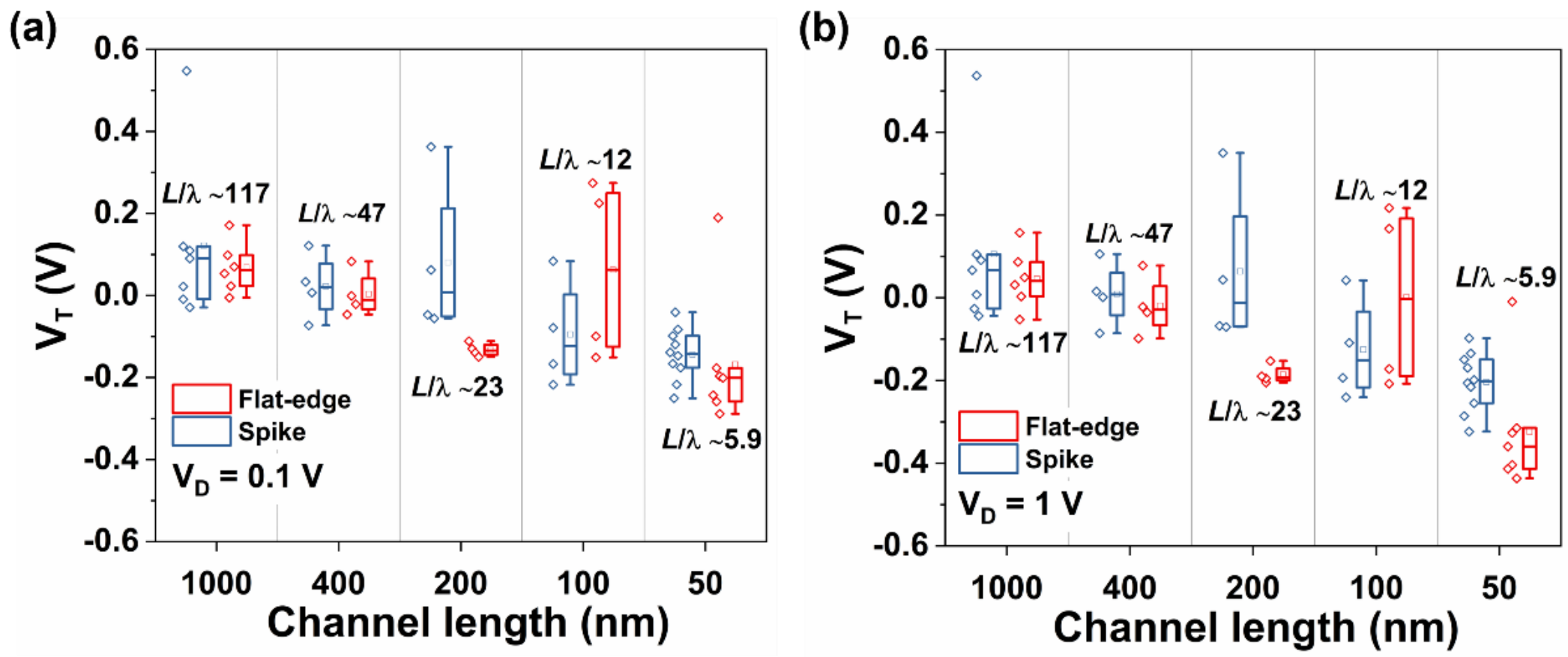

**Figure S5**. $V_T$ comparison between nanospike and flat-edge devices at various $L_{ch}$

**Supporting Information 6**

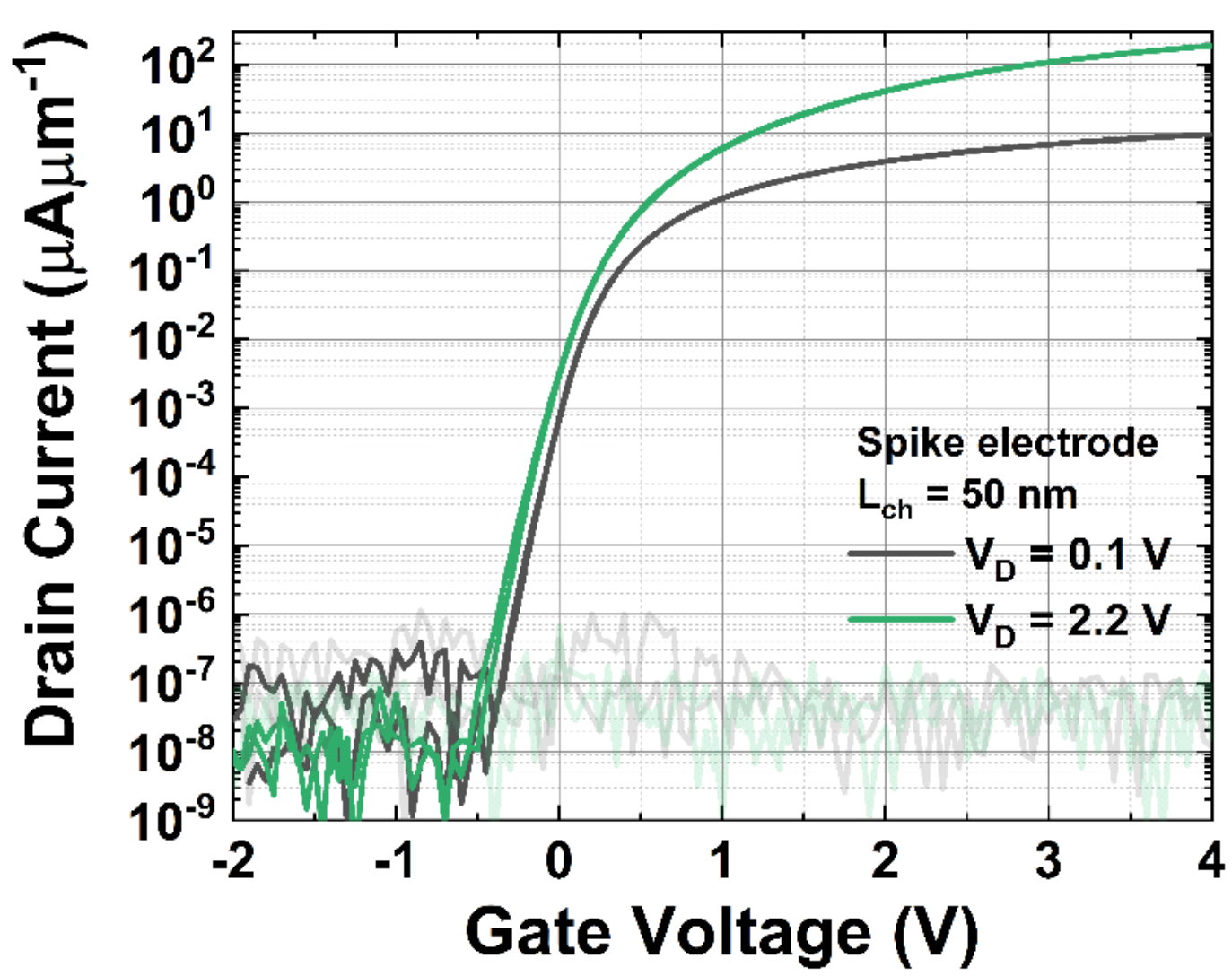

**Figure S6**. (a) $I_D$-$V_G$ transfer characteristics of 50 nm $L_{ch}$ nanospike device at $V_D$= 0.1 V and 2.2 V. SS remains low at 127 mV/dec under $V_D$ = 2.2 V.

**Supporting Information 7**

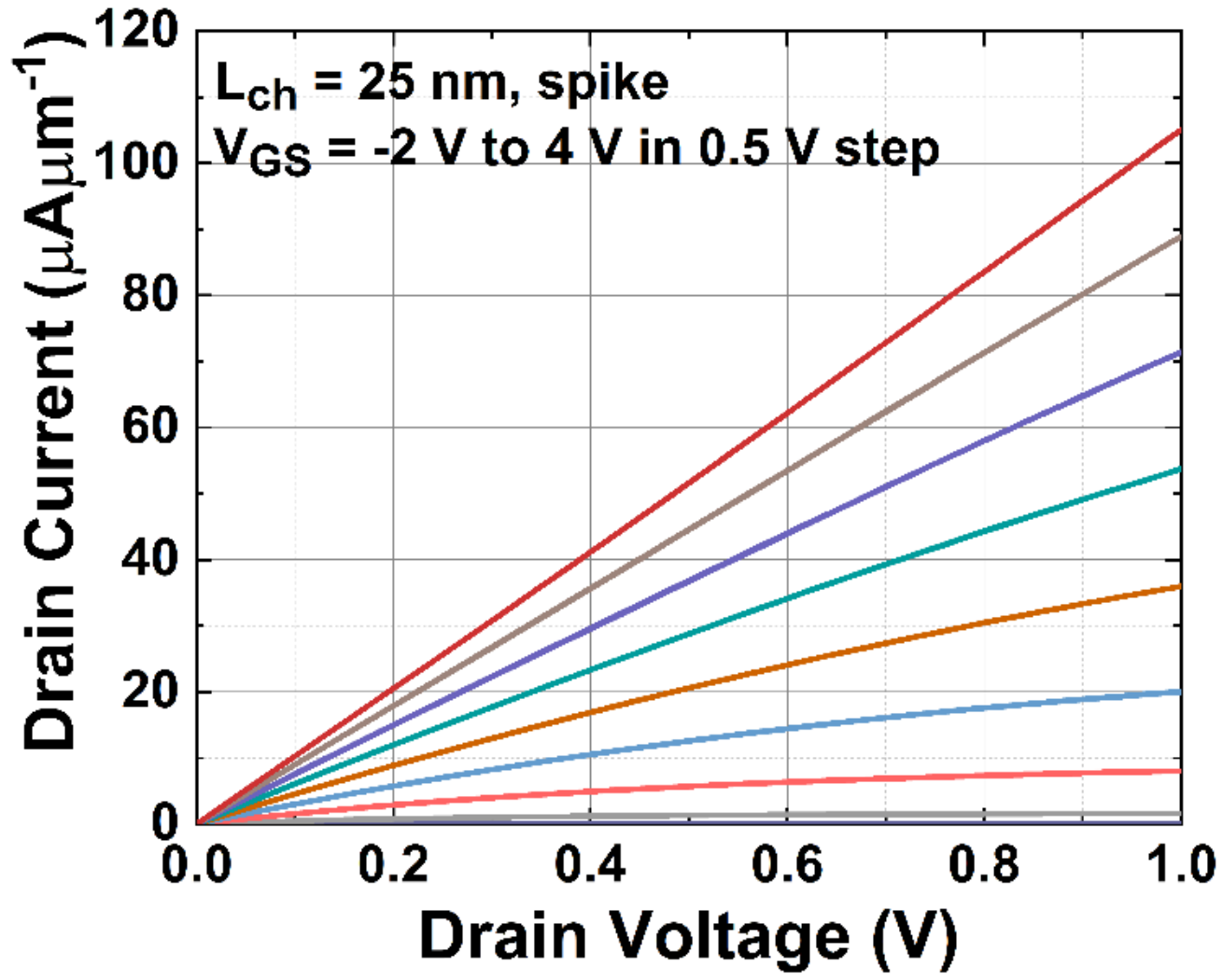

**Figure S7**. $I_D$-$V_D$ output characteristics of 25 nm $L_{ch}$ nanospike device.

**Supporting Information Table 1**

Performance of sub-50 nm $L_{ch}$ devices with different oxide semiconductors

| REF | Channel | Structure | Gate dielectric | $T_{ch}$ | $L_{ch}$ | $L_{ch}/\lambda$ | $I_{on}/I_{off}$ | SS [mV/dec] | DIBL [mV/V] |
|---|---|---|---|---|---|---|---|---|---|
| This work | IGZO | BG (spike) | $Al_2O_3$ (9 nm) | 6 nm (sputter) | 50 nm | - | $10^9$ | 130 | 55 |
| This work | IGZO | BG (spike) | $Al_2O_3$ (9 nm) | 6 nm (sputter) | 25 nm | - | $10^9$ | 160 | 88 |
| This work | IGZO | BG (flat-edge) | $Al_2O_3$ (9 nm) | 6 nm (sputter) | 50 nm | 5.8 | $10^9$ | 142 | 161 |
| [1] | IGZO | BG | $HfO_2$ (6nm) | 1.5 nm (ALD) | 60 nm | 24.5 | $10^{11}$ | 68 | 30 |
| [2] | IGZO | BG | $HfO_2$ (15 nm) | 8 nm (sputter) | 45 nm | 5 | $10^8$ | 105 | 120 |
| [3] | IGZO | BG | $HFO_2$ (10 nm) | 3.6 nm (sputter) | 38 nm | 7.8 | $10^7$ | 74 | 187 |
| [4] | CAAC-IGZO | Trench-gate-self-aligned | 11 nm (E.O.T) | 15 nm (sputter) | 27 nm | - | $10^8$ | 97 | 120 |
| [5] | $In_2O_3$ | BG | $HfO_2$ (3 nm) | 3.5 nm (ALD) | 7 nm | 2.6 | 900 | 540 | 2175 |
| [6] | $In_2O_3$ | BG | $HfO_2$ (E.O.T 0.84 nm) | 2.5 nm (ALD) | 8 nm | 3.6 | 37 | 1100 | 1400 |
| [7] | ITO | BG | HfLaO (5 nm) | 3.5 nm (sputter) | 10 nm | 6.6 | $10^{10}$ | 176 | 488 |

*BG = Bottom gate

*GAAC-IGZO = c-axis aligned crystalline-IGZO

Relative permittivity of $HfO_2$ ~15 and IGZO, $In_2O_3$, ITO ~10 are used for calculating $\lambda$, if not mentioned in each manuscript.

**Supporting Information 8**

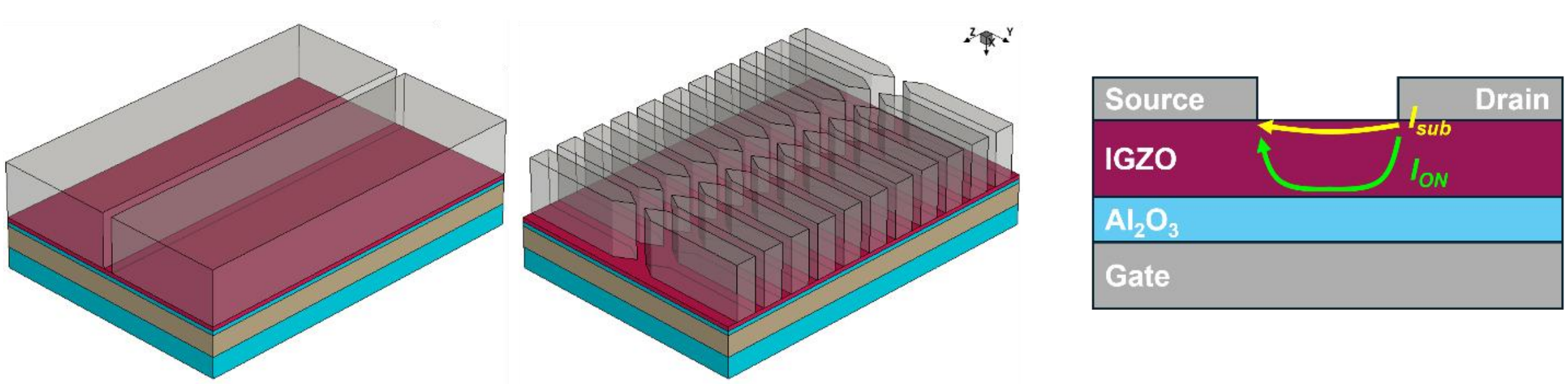

**Figure S8**. TCAD simulation structures for (a) flat-edge and (b) nanospike electrode devices. The channel width is set to 0.34 µm, and the number of fingers for nanospike device is 11. The channel

and gate oxide thickness are both 4 nm to represent highly scaled devices. (c) Schematic illustration of primary conduction for the subthreshold and on-state regime.

**Supporting Information 9**

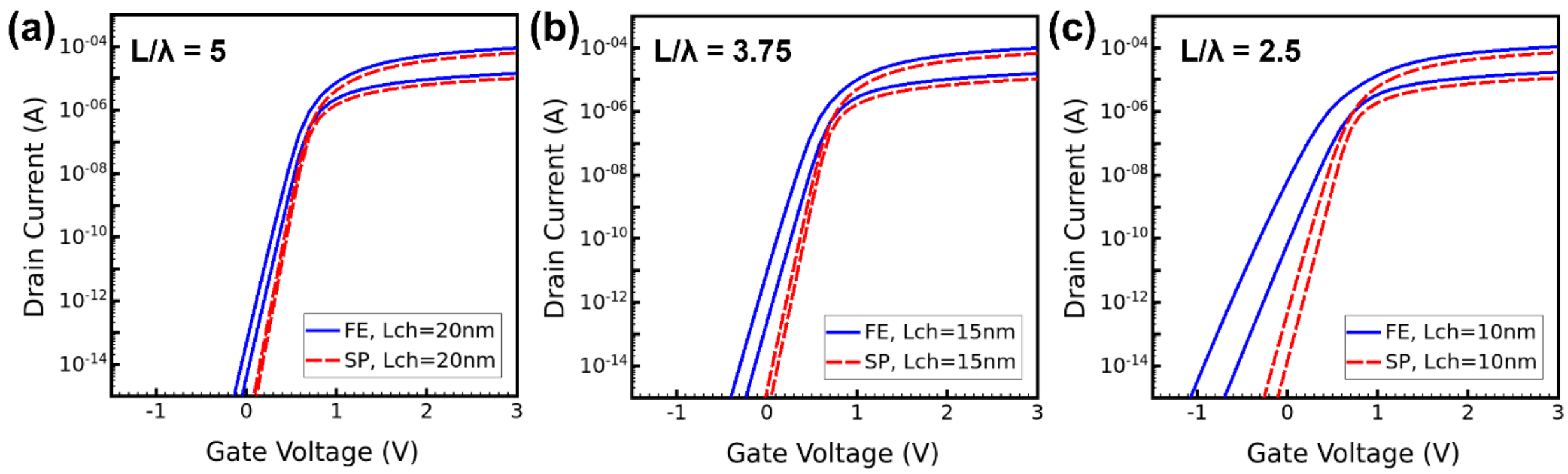

**Figure S9**. Comparisons of $I_D$-$V_G$ transfer curve between flat-edge (FE) and nanospike (SP) devices for $L_{ch}$ of (a) 20 nm, (b) 15 nm and (c) 10 nm. The gate oxide thickness ($T_{ox}$) and IGZO thickness are ($T_{IGZO}$) both 4 nm (k = 10) to represent highly scaled devices. $V_{DS}$ is set to 0.1 and 0.7 V for low $V_D$ and high $V_D$, respectively.

**Supporting Information 10**

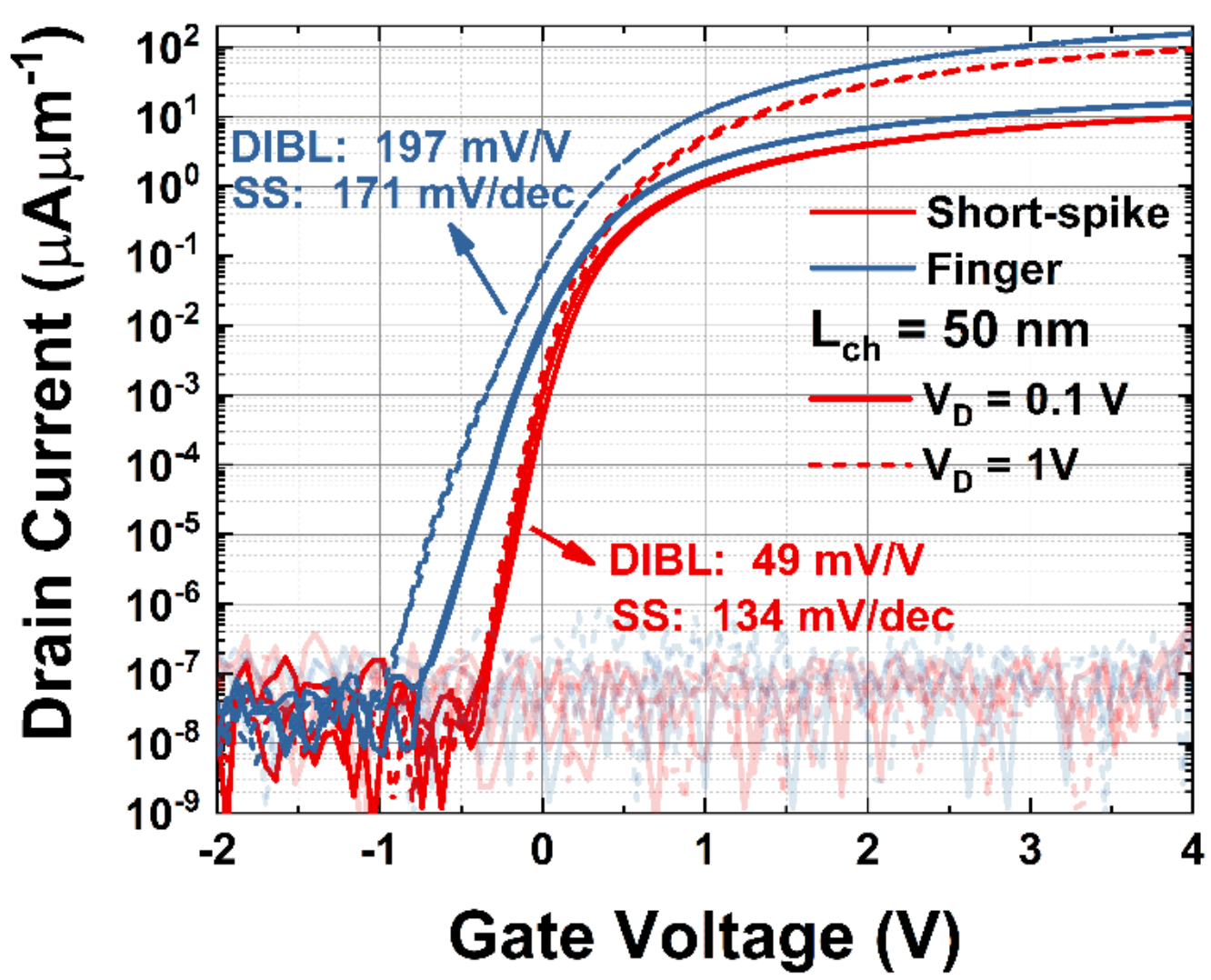

**Figure S10**. $I_D$-$V_G$ transfer characteristics of 50 nm $L_{ch}$ finger and short-nanospike device.

## Supporting Information 11

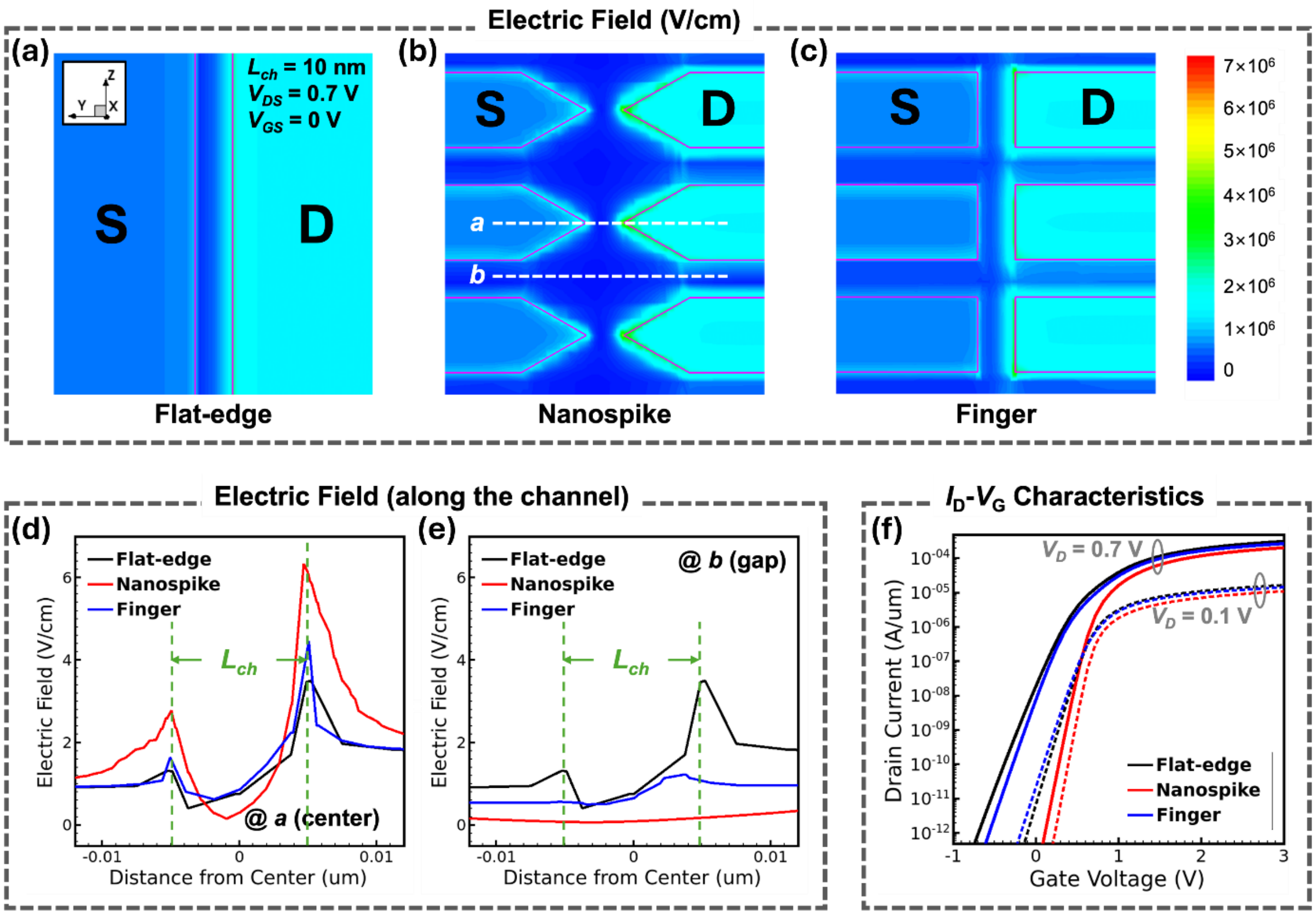

**Figure S11.** Electric field distributions of (a) flat-edge, (b) nanospike and (c) finger electrode devices in the subthreshold regime, observed at the top surface of IGZO channel. (d, e) Electric field profiles along cut-lines a and b at the top surface of IGZO channel (as defined in Fig. **S11b**) (f) Comparisons of $I_D$-$V_G$ transfer curve between flat-edge, nanospike and finger devices.

Based on TCAD simulations results, the finger electrode is observed to be less effective in mitigating SCEs, consistent with the experimental data reported in main MS. The finger electrode device exhibits a similar electric field distribution along the channel to that of the flat-edge device in the subthreshold regime, which accounts for its limited effectiveness in suppressing SCEs.

**Supporting Information 12**

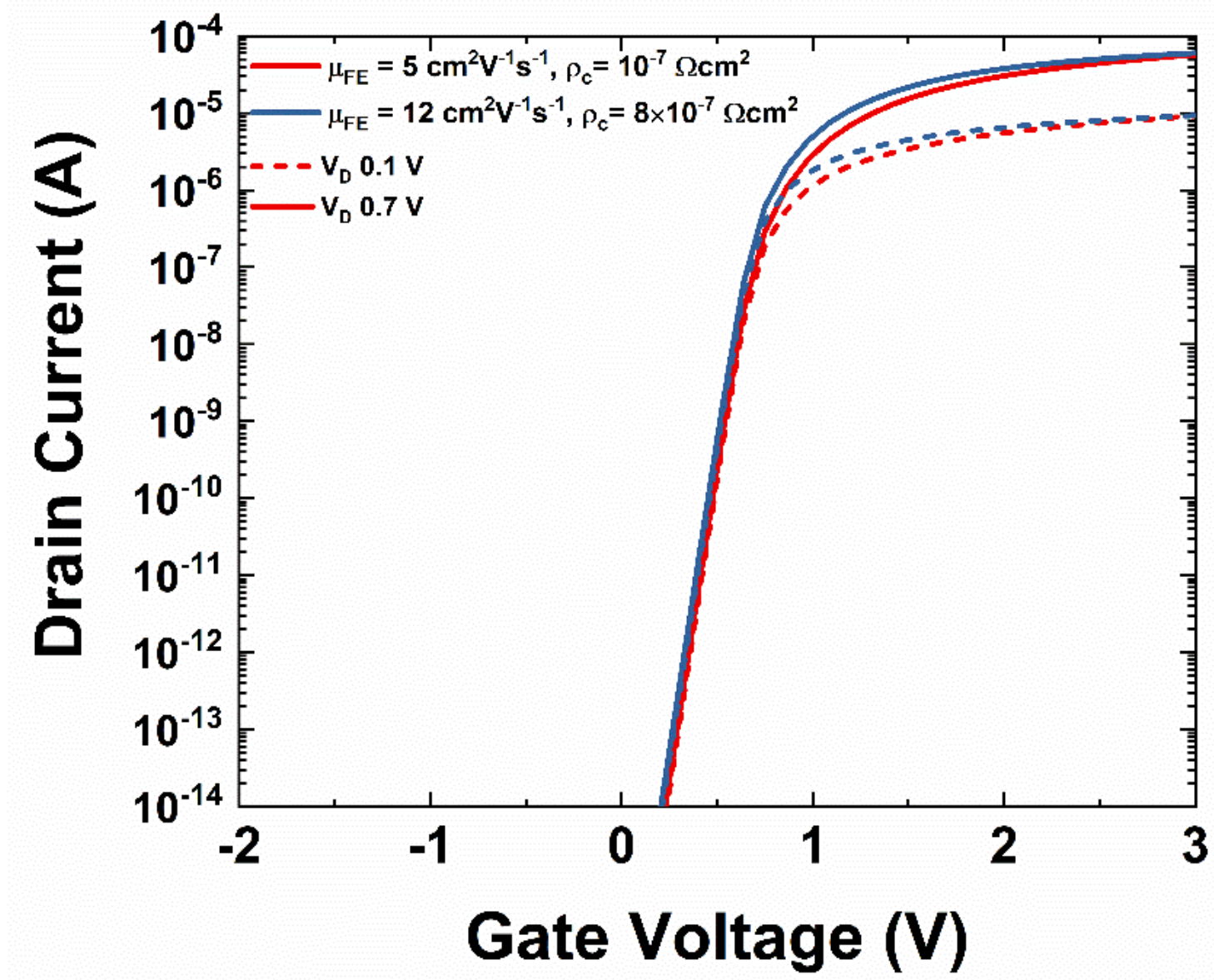

**Figure S12**. Simulated transfer characteristics of a 50 nm channel FET with specific contact resistivities of $1\times10^{-7}$ and $8\times10^{-7}$ Ohm-cm$^2$ at $V_D = 0.1$ V and 0.7 V. Increasing the specific contact resistivity from $1\times10^{-7}$ to $8\times10^{-7}$ cm$^2$V$^{-1}$s$^{-1}$ requires the maximum mobility to be increased from 5 cm$^2$/V-s to 12 cm$^2$V$^{-1}$s$^{-1}$ for the drain currents to match in the on-state. This suggests that contact resistance effects are very important at these channel lengths